\documentclass[epj]{webofc}
\usepackage[varg]{txfonts}   
\woctitle{MESON2018 - the 15$^\textrm{th}$ International Workshop on Meson Physics}
\begin{document}
\selectlanguage{english}
\title{Recent progress in the partial-wave analysis of the diffractively produced $\pi^-\pi^+\pi^-$ final state at \textsc{Compass}}

\author{Fabian~Krinner\inst{1}\fnsep\thanks{\email{fabian-krinner@mytum.de}} 
       \\for the \textsc{Compass} Collaboration
}

\institute{Technische Universit\"at M\"unchen, Munich, Germany
          }

\abstract{
  The \textsc{Compass} spectrometer at CERN has collected a large data set for 
  diffractive three-pion production of $46\times10^6$ exclusive events. Based on
  previous conventional Partial-Wave Analyses (PWA), we performed a ``freed-isobar
  PWA'' on the same data, removing model assumptions on the dynamic isobar
  amplitudes for dominating waves. In this analysis, we encountered continuous
  mathematical  ambiguities, which we were able to identify and resolve. This
  analysis gives an unprecedented insight in the interplay of $2\pi$ and $3\pi$
  dynamics in the process. As an example we show results for a spin-exotic wave
  $J^{PC}_{X^-}=1^{-+}$ wave.
}
\maketitle
\section{Diffractive $\mathbf{3\pi}$ production at \textsc{Compass}}\label{sec::data}

In this work, we perform a Partial-Wave Analysis (PWA) of the diffractive process 
\[
  \pi_\text{beam}^- + p_\text{target} \to X^- + p_\text{recoil}  \to \pi^-\pi^+\pi^- + p_\text{recoil} 
\]
measured by the \textsc{Compass} experiment at CERN using an $190\,\text{GeV}/c$ 
negative hadron beam---consisting to 97\% of negative pions---impinging on a 
liquid hydrogen target. For this process the \textsc{Compass} collaboration has 
recorded a very large data set of $46\times10^6$ exclusive events, performed a 
very detailed PWA on this data using a set of 88 partial waves 
\cite{massIndepPaper}, and extracted resonance masses and widths of eleven 
intermediate isovector resonances $X^-$ \cite{massDepPaper}.

\section{Freed-isobar PWA}\label{sec::freedisobar}

The PWA preformed in Ref.~\cite{massIndepPaper} relies on the isobar model, in 
which the decay of the intermediate state $X^-$ into the $3\pi$ final state is 
modeled as a sequence of two-particle decays involving a second intermediate 
state $\xi^{\,0}$, the isobar:
\[
 X^-\to\xi^{\,0}+\pi^-_\text{bachelor} \to \pi^-\pi^+\pi^-
;\]
the best known examples for such isobars are the $\rho(770)$, $f_0(980)$, and 
$f_2(1270)$ resonances. In the conventional approach, the dynamic amplitudes of 
these isobar resonances $\Delta_{\xi^{\,0}}\big(m_{\xi^{\,0}}\big)$---often 
called ``line shape''---are a necessary input for the PWA model. The most common
 example for such a dynamic isobar amplitude is the well-known Breit-Wigner 
amplitude with given resonance mass and width. However, the necessity for fixed 
dynamic isobar amplitudes in a conventional PWA hast several disadvantages, 
since
\begin{itemize}
\item it is not a priori clear, which isobar resonances to include in the 
      analysis model, 
\item Breit-Wigner amplitudes may not give an accurate description of all 
      isobars,
\item and overlapping Breit-Wigner amplitudes violate theoretical requirements.
\end{itemize}
To avoid these drawbacks of the conventional approach, we use an analysis 
technique called ``freed-isobar PWA''---also often named ``model-independent 
PWA''---where we replace the fixed dynamic isobar amplitudes by sets of bin-wise 
constant functions:
\begin{equation}\label{eq::repl}
 \Delta_{\xi^{\,0}}\big(m_{\xi^{\,0}}\big)\to\sum_\text{bins}\Delta_{\xi^{\,0}}^\text{bin}\big(m_{\xi^{\,0}}\big)
 \quad\text{with}\quad \Delta_{\xi^{\,0}}^\text{bin}\big(m_{\xi^{\,0}}\big)=\begin{cases}1 & \text{if $m_{\xi^{\,0}}$ in the bin},\\0 & \text{otherwise.}\end{cases}
\end{equation}
Since every single $m_{\xi^{\,0}}$ bin behaves like an independent partial wave 
in the analysis model the freed-isobar approach allows to re-used the existing 
analysis scheme with a much higher number of degrees of freedom. Bin-wise 
approximations to the dynamic isobar amplitudes are hereby encoded in the 
strengths and relative phases of these individual partial waves. This approach 
allows to resolve the process in terms of the angular-momentum quantum numbers 
and the mass of the isobar. Partial waves with dynamic isobar amplitudes 
replaced this way will be called ``freed'' from hereon.

We performed such a freed-isobar analysis on the data set introduced in 
Sect.~\ref{sec::data} using the same wave-set and freeing the following 12 
of the total 88 waves:
\begin{align*}
 0^{-+}&0^+[\pi\pi]_{0^{++}}\pi \text{S}, &\ 1^{++}&0^+[\pi\pi]_{0^{++}}\pi \text{P}, &\ 2^{-+}&0^+[\pi\pi]_{0^{++}}\pi \text{D}, &\ 2^{-+}&0^+[\pi\pi]_{2^{++}}\pi \text{S}, \\
 0^{-+}&0^+[\pi\pi]_{1^{--}}\pi \text{P}, &\ 1^{++}&0^+[\pi\pi]_{1^{--}}\pi \text{S}, &\ 2^{-+}&0^+[\pi\pi]_{1^{--}}\pi \text{P}, &\ 2^{-+}&1^+[\pi\pi]_{1^{--}}\pi \text{P}, \\
 1^{-+}&1^+[\pi\pi]_{1^{--}}\pi \text{P}, &\ 1^{++}&1^+[\pi\pi]_{1^{--}}\pi \text{S}, &\ 2^{-+}&0^+[\pi\pi]_{1^{--}}\pi \text{F}, &\ 2^{++}&1^+[\pi\pi]_{1^{--}}\pi \text{D}.
\end{align*}
The freed waves were chosen to be the 11 waves with the highest intensity in the
 conventional analysis plus the spin-exotic $J^{PC}_{X^-}=1^{-+}$ wave, which is
 a wave of major interest. Since partial waves with identical angular-momentum 
quantum numbers are absorbed in a single freed wave, this leaves 72 waves with 
fixed dynamic isobar amplitudes in the model.

The analysis was performed in 50 independent bins in the invariant mass of the 
$3\pi$ system, from $0.5$ to $2.5\,\text{GeV}/c^2$ and four non-equidistant bins
 in the four-momentum transfer $t^\prime$ in the analyzed region from $0.1$ to 
$1.0\,(\text{GeV}/c)^2$, giving a total of 200 independent fits. The width of 
the $m_{\xi^{\,0}}$ bins was chosen to be $40\,\text{MeV}/c^2$, with smaller 
widths in the regions of known resonances: $20\,\text{MeV}/c^2$ in the regions 
of the $\rho(770)$ and the $f_2(1270)$, and $10\,\text{MeV}/c^2$ in the region 
of the $f_0(980)$.

\section{Zero modes in the freed-isobar analysis}\label{sec::zeroMode}

The fact, that models in a freed-isobar PWA have a much higher number of degrees 
of freedom, may lead to the appearance of continuous ambiguities, caused by 
exact cancellations between different terms of the amplitude. Such 
cancellations, which we call ``zero modes'' from hereon, therefore have to be 
identified and the corresponding ambiguities have to be resolved. 

Since this article focuses on the spin-exotic wave, we show how a zero mode 
arises within this particular wave. The dependence on the kinematic variables of
 the decay $X^-\to\pi_1^-\pi_2^+\pi_3^-$ of the spin-exotic wave is determined 
by its angular-momentum quantum numbers and given by:
\begin{equation}
\hat A_{1^{-+}}^{(12)}\propto(\vec p_1\times\vec p_3)\Delta_{1^{-+}}\big(m^{(12)}_{\xi^{\,0}}\big)
,\end{equation}
where the appearing three-momenta are defined in a rest system of $X^-$ and we 
have assumed that the isobar is formed by $\pi_1^-$ and $\pi_2^+$. However, 
since there are two identical $\pi^-$ in the final-state, the amplitude has to 
be Bose symmetrized and the total amplitude of the spin-exotic wave is:
\begin{equation}\label{eq::zeroModeSee}
A_{1^{-+}}=\hat A_{1^{-+}}^{(12)}+\hat A_{1^{-+}}^{(23)}\propto(\vec p_1\times\vec p_3)\left[\Delta_{1^{-+}}\big(m^{(12)}_{\xi^{\,0}}\big)-\Delta_{1^{-+}}\big(m^{(23)}_{\xi^{\,0}}\big)\right]
,\end{equation}
where the respective minus sign stems from the exchange of $\pi^-_1$ and 
$\pi^-_3$ in Bose symmetrization and the antisymmetry of the cross product.
From this equation, we can easily see, that the amplitude of the spin-exotic 
wave is invariant under a change of the dynamic isobar amplitude by:
\begin{equation}
\Delta_{1^{-+}}\big(m_{\xi^{\,0}}\big)\to\Delta_{1^{-+}}\big(m_{\xi^{\,0}}\big) + \mathcal{C}
,\end{equation}
since both Bose-symmetrization terms exactly cancel in 
Eq.~(\ref{eq::zeroModeSee}). Therefore, also the intensity and the likelihood 
function are invariant under this transformation. Thus, we have identified a 
zero-mode in the spin-exotic wave, where the corresponding ambiguity is encoded 
by the complex-valued coefficient $\mathcal{C}$.

Since the likelihood function is invariant under a change of the zero-mode 
coefficient $\mathcal{C}$, the fitting algorithm may find a solution with any 
arbitrary value for it, which might not represent the physical one. Therefore, 
we have to adjust $\mathcal{C}$ in a second fit step, using additional 
conditions on the resulting dynamic isobar amplitude as constraint. In the 
case of the spin-exotic wave, we required the resulting dynamic isobar 
amplitude to be as close as possible to a Breit-Wigner shape for the dominating 
$\rho(770)$ resonance within the scope of the sole parameter $\mathcal{C}$. To 
minimize possible effects of excited $\rho^\prime$ resonances, we limited the 
fit range to isobar masses below $1.12\,\text{GeV}/c^2$.

Note, that this second fit step fixes only a single complex-valued degree of 
freedom, while $n_\text{bins}-1$ degrees of freedom still remain free. We 
validated this method in several Monte-Carlo studies and especially verified, 
that one cannot create resonance signals with arbitrary parameters in the 
process. More information on zero-mode ambiguities and their resolution can be 
found in Ref.~\cite{zeroModePaper}.

\section{Results of the freed-isobar analysis}

With the model defined in Sec.~\ref{sec::freedisobar} and the method to resolve 
the zero-mode ambiguity given in the previous section, we can analyze the 
data-set of Sec.~\ref{sec::data} and obtain an unprecedented insight into the 
dynamics of diffractive three-pion production.

\begin{figure}[h]
\centering
\includegraphics[width=6.cm,clip]{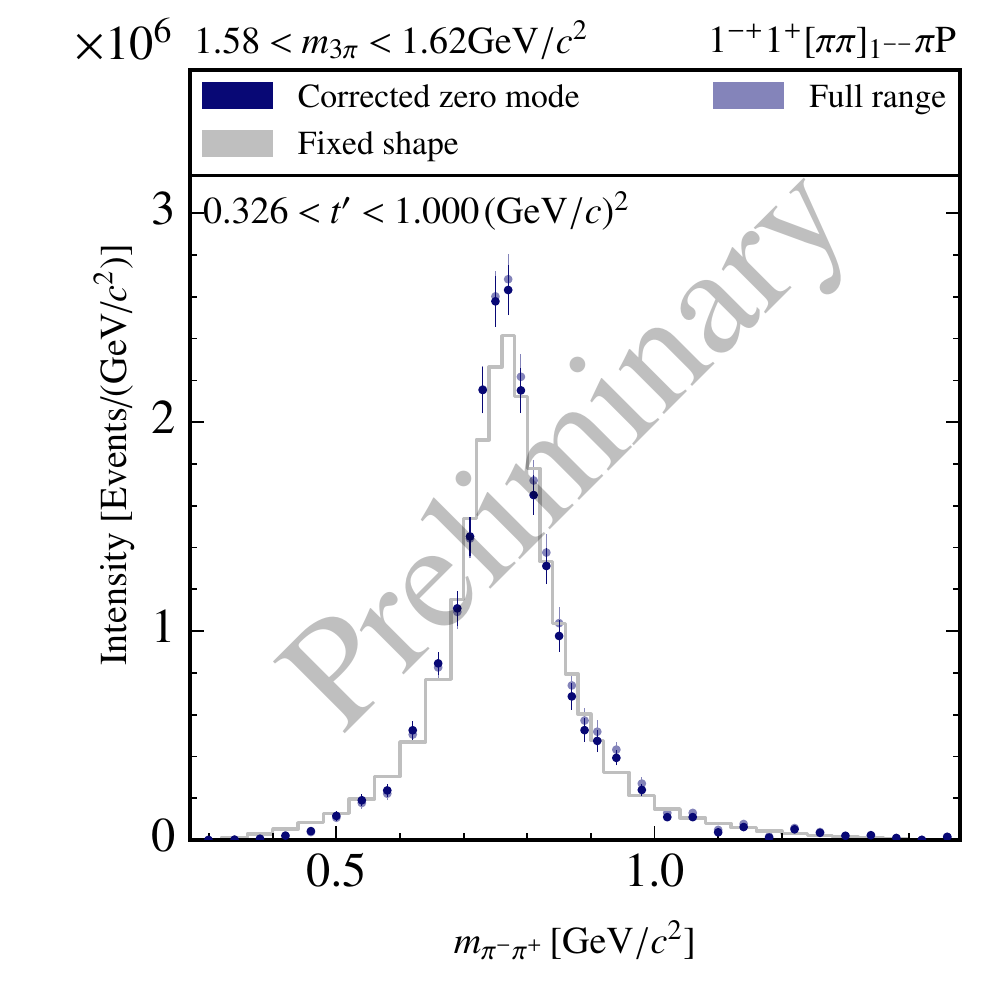}
\includegraphics[width=6.cm,clip]{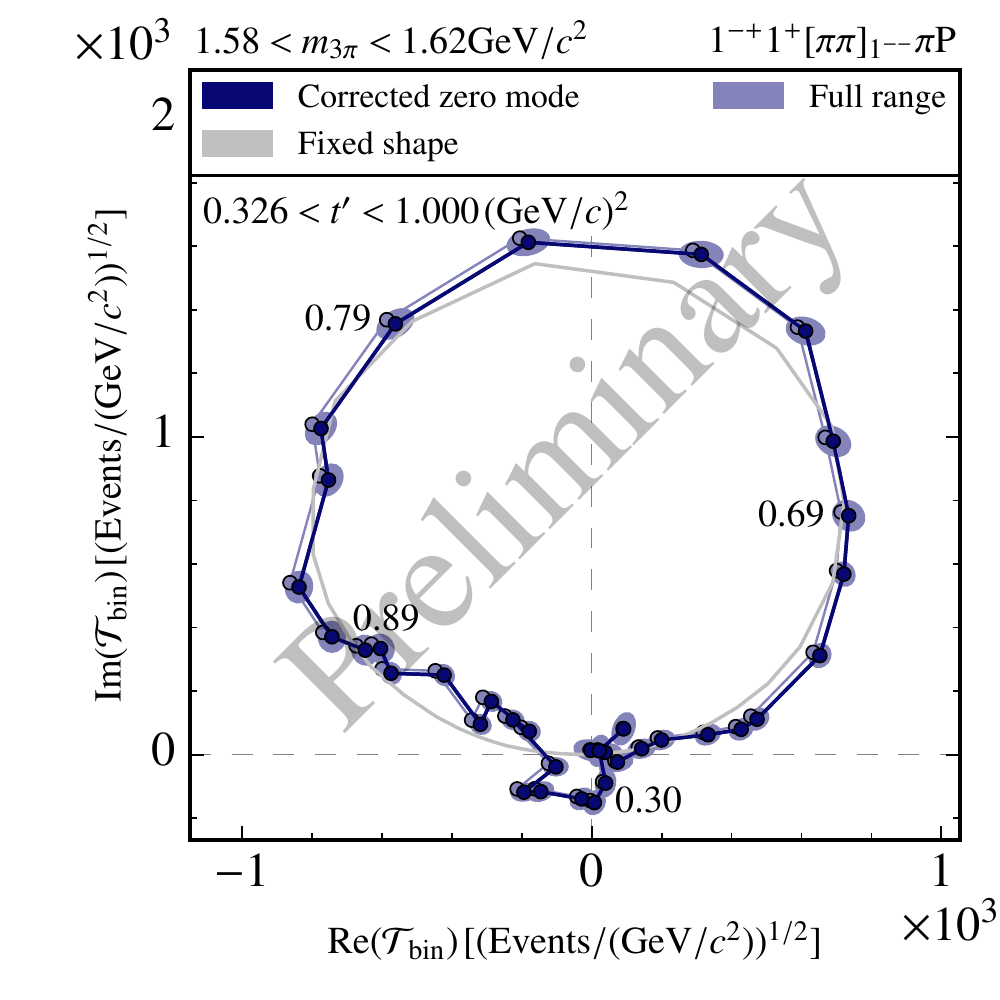}
\caption{Results for the dynamic isobar amplitude of the spin-exotic wave for 
         a single bin in $m_{3\pi}$ and $t^\prime$ in two different 
         representations. The resulting intensity distribution is shown on the 
         left, and the Argand diagram on the right. The gray line indicates the 
         constraint used to resolve the zero-mode ambiguity.}
\label{fig-1}       
\end{figure}

The result for a single $(m_{3\pi},t^\prime)$ bin is shown in Fig.\ref{fig-1}, 
we see, that the resulting dynamic isobar amplitude is dominated by the 
$\rho(770)$ resonance, as expected. The fixed Breit-Wigner amplitude is a good 
approximation to the dynamic isobar amplitude, with some significant deviations,
 especially in the peak region. Such deviations might be caused by non-resonant 
contributions to the process or re-scattering effects with the third pion.

\begin{figure}[h]
\centering
\includegraphics[width=5.3333333333cm,clip]{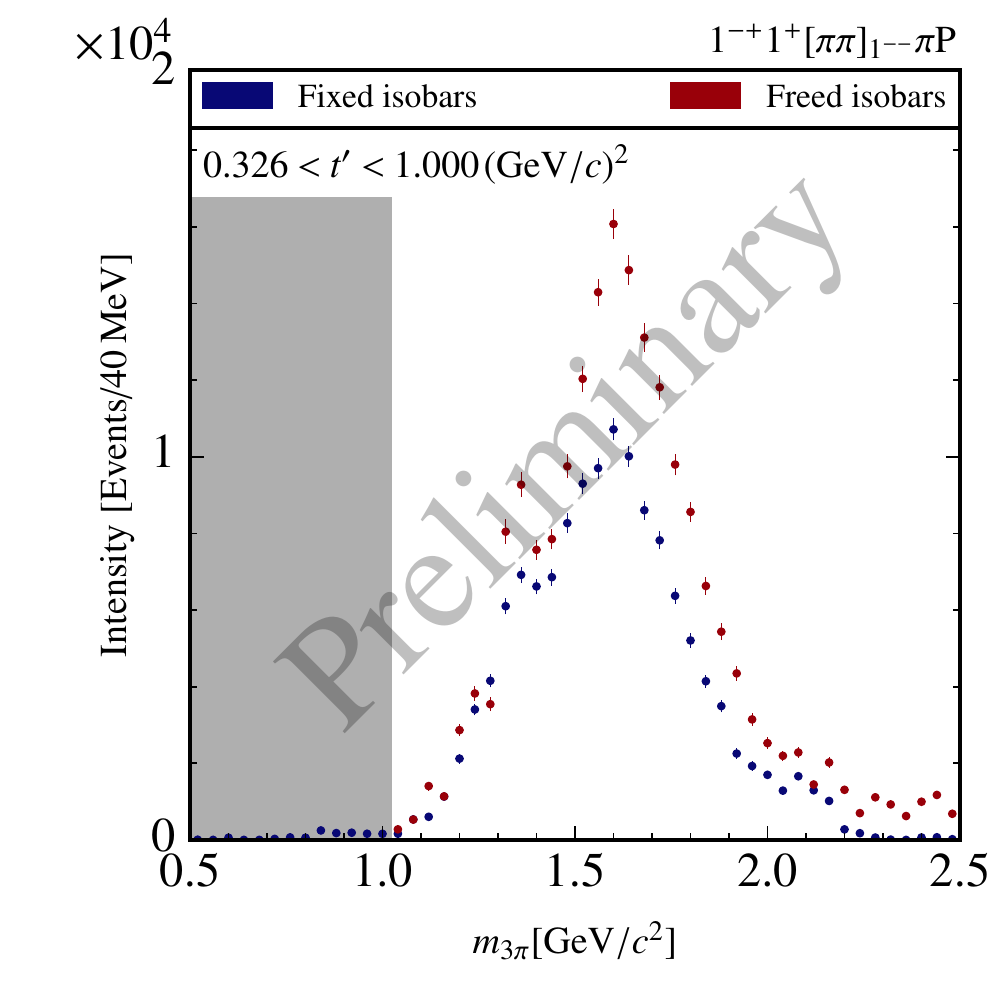}
\includegraphics[width=6.6666666666cm,clip]{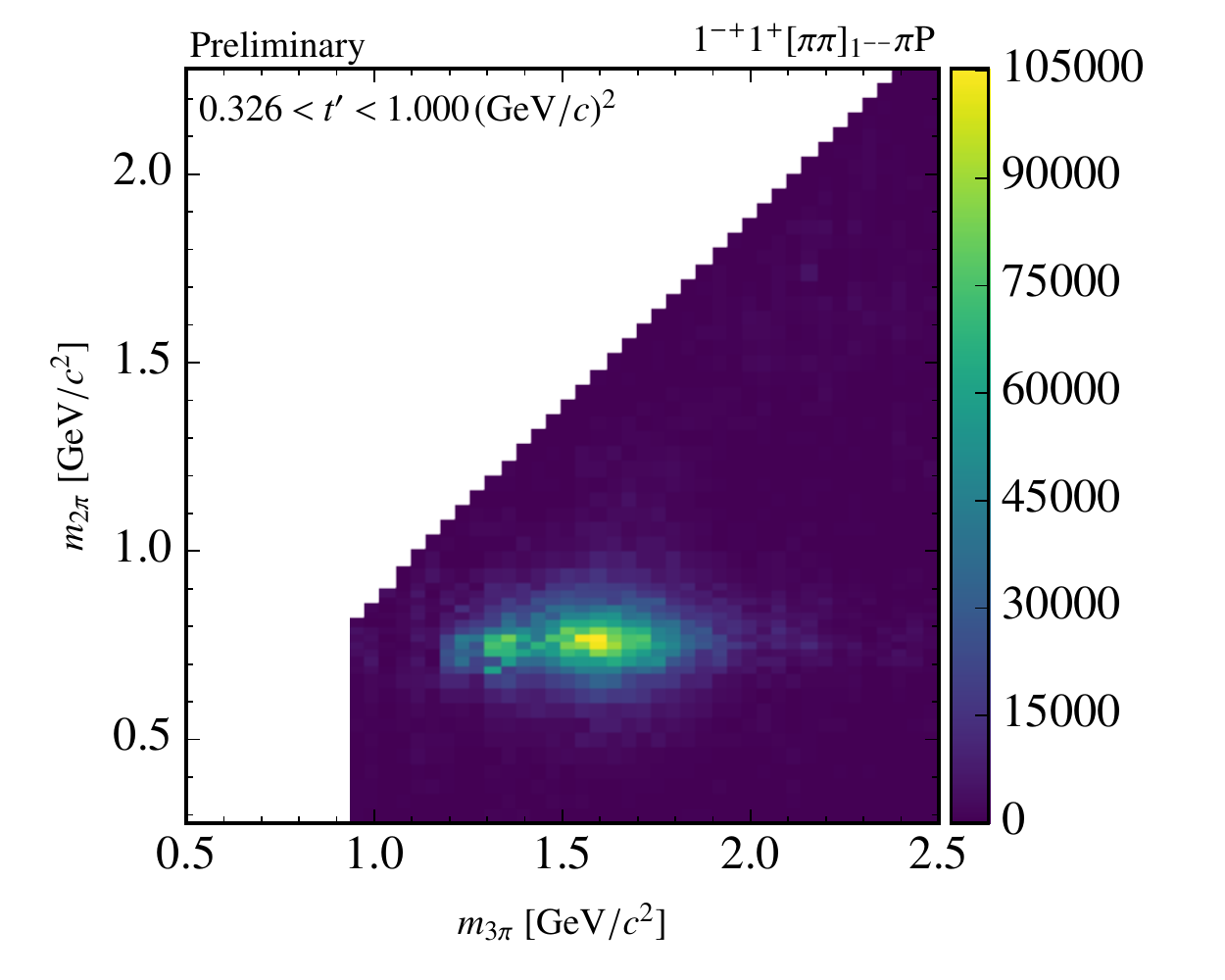}
\caption{Results of the freed-isobar analysis of the spin-exotic wave as a 
         function of $m_{3\pi}$. The two-dimensional intensity distribution is 
         shown on the right and the coherent sum of all $m_{\xi^{\,0}}$ bins on 
         the left, compared to the results of the conventional PWA.}
\label{fig-2}       
\end{figure}

Looking at the dependence of the freed-isobar results on the $m_{3\pi}$ bin, 
as shown in Fig.~\ref{fig-2}, we find a nice correlation of the intensity 
distributions in $m_{3\pi}$ and $m_{\xi^{\,0}}$, which corresponds to the 
dominating decay $\pi_1^-(1600)\to\rho(770)\pi^-$. The coherent sum of all 
$m_{\xi^{\,0}}$ has a similar peak position and width, as the result of the 
conventional analysis in Ref.~\cite{massIndepPaper}, but a higher total 
intensity. This shows that the $\pi_1^-(1600)$ resonance found in the 
conventional analysis of Ref.~\cite{massDepPaper} is not an artifact of the 
fixed dynamic isobar amplitudes used.

\section{Conclusions}

We performed an extended freed-isobar PWA for diffractive $3\pi$ production, for
 which the \textsc{Compass} spectrometer has collected large exclusive data set 
of $46\times10^6$ events. In this ana\-ly\-sis, we encountered continuous 
mathematical ambiguities---zero modes---, which are identified and resolved. The
 results for the spin-exotic wave showed, that the dynamic isobar amplitude is 
dominated by the $\rho(770)$ resonance, as expected. However, some small but 
significant deviations from a pure Breit-Wigner shape are visible. We compare 
our findings to the conventional PWA method and find a peak compatible with the 
$\pi_1^-(1600)$ resonance of Refs.~\cite{massIndepPaper,massDepPaper}.


\begin{thebibliography}{00}
%
%

\bibitem{massIndepPaper}
\textsc{Compass} collaboration (C.~Adolph~{\it et al.}), 
Phys.Rev. {\bf D95} (2017) no.3, 032004

\bibitem{massDepPaper}
\textsc{Compass} collaboration (R.~Akhunzyanov~{\it et al.}), 
arXiv:1802.05913 [hep-ex] (2018)

\bibitem{zeroModePaper}
F.~Krinner, D.~Greenwald, D.~Ryabchikov, B.~Grube, and S.~Paul,\\ 
Phys. Rev. {\bf D97} (2018), 114008

\end{thebibliography}
\end{document}